\documentclass[
superscriptaddress,
amsmath,amssymb,
aps,
prl,
showkeys
]{revtex4-2}

\usepackage{graphicx}
\graphicspath{{./figures/}}
\usepackage[dvipsnames]{xcolor}
\usepackage{ulem}
\usepackage{dcolumn}
\usepackage{bm}
\usepackage{hyperref}
\hypersetup{colorlinks, linkcolor = [rgb]{0, 0, 0.5}, citecolor = [rgb]{0,0.0,0.5}, urlcolor = [rgb]{0,0.0,0.5}}
\usepackage[caption=false]{subfig}
\usepackage{siunitx}
\usepackage[capitalise]{cleveref}

\allowdisplaybreaks

\begin{document}

\title{The Sivers asymmetry of vector meson production in semi-inclusive deep inelastic scattering}

\newcommand*{\SDU}{Key Laboratory of Particle Physics and Particle Irradiation (MOE), Institute of Frontier and Interdisciplinary Science, Shandong University, Qingdao, Shandong 266237, China}\affiliation{\SDU}
\newcommand*{\SCNT}{Southern Center for Nuclear-Science Theory (SCNT), Institute of Modern Physics, Chinese Academy of Sciences, Huizhou 516000, China}\affiliation{\SCNT}

\author{Yongjie Deng}\affiliation{\SDU}
\author{Tianbo~Liu}\email{liutb@sdu.edu.cn}\affiliation{\SDU}\affiliation{\SCNT}
\author{Ya-jin~Zhou}\email{zhouyj@sdu.edu.cn}\affiliation{\SDU}

\begin{abstract}
The transverse single-spin asymmetry for $\rho^0$ production in semi-inclusive deep inelastic scattering was recently reported by the COMPASS Collaboration. 
Using the Sivers functions extracted from pion and kaon productions, we perform a calculation of the Sivers asymmetry within the transverse momentum dependent factorization. 
Our results are consistent with the COMPASS data, supporting the universality of the Sivers functions in the SIDIS process for different final state hadrons within current experimental uncertainties. 
While different parametrizations of the Sivers function from global analyses allow describing the data equally well, 
we obtain very different predictions on the Sivers asymmetry of $\rho$ and $K^*$ productions at electron-ion colliders, which therefore are expected to provide further constraints.  
\end{abstract}


\maketitle

\section{Introduction}
\label{sec: introduction}

A key issue in modern nuclear and particle physics is to understand the internal structure of the nucleon at the level of quarks and gluons, the fundamental degrees of freedom of quantum chromodynamics (QCD). Due to the nonperturbative nature of QCD at hadronic energy scales, evaluating nucleon structures from first principles remain a challenging task, although much progress has been made in recent years~\cite{Ji:2013dva, Radyushkin:2017cyf, Ma:2014jla}. 
As quarks and gluons cannot be directly observed, known as the color confinement, the connection between experimental results and theoretical calculations is primarily made through the QCD factorization.

The semi-inclusive deep inelastic scattering (SIDIS) is one of the primary processes for investigating the three-dimensional partonic structures of the nucleon.
In this reaction a high-energy lepton beam scatters off a target nucleon and a specified hadron along with the scattered lepton is detected in the final state.
At low transverse momentum the SIDIS cross section can be described as the convolution of the transverse momentum dependent (TMD) parton distributions (PDFs), the TMD fragmentation functions (FFs), and the short-distance hard part.
Taking into account the spin of the parton and the nucleon, one can also investigate their correlation with the transverse motion of the parton, unraveling rich structures of the nucleon. The polarized TMD PDFs and FFs will lead to azimuthal modulations in polarized SIDIS process. Among them, we focus on the target transverse single spin asymmetry (TSSA), referred to as the Sivers asymmetry, which has been extensively studied in recent decades. 
Within the TMD factorization, the Sivers asymmetry can arise from the convolution of the Sivers distribution function $f_{1T}^\perp(x,k_\perp^2)$, where $x$ and $k_\perp$ are the longitudinal momentum fraction and the transverse momentum carried by the parton with respect to the parent nucleon, and the unpolarized TMD FF $D_1(z,p_\perp^2)$, where $z$ and $p_\perp$ are the longitudinal momentum fraction and transverse momentum of the final-state hadron with respect to the fragmenting parton.
The Sivers function, as one of the leading-twist TMD PDFs,
characterizes the correlation between the transverse momentum of the parton and the transverse spin of the nucleon.

The Sivers asymmetry in SIDIS was first measured and found to be sizable by the HERMES Collaboration with charged pion productions~\cite{HERMES:2004mhh}. This effect was further confirmed by subsequent HERMES experiments~\cite{HERMES:2009lmz, HERMES:2020ifk}, as well as by the COMPASS~\cite{COMPASS:2005csq, COMPASS:2016led, COMPASS:2008isr, COMPASS:2012dmt, COMPASS:2010hbb, COMPASS:2006mkl, COMPASS:2018ofp}, and Jefferson Lab (JLab)~\cite{JeffersonLabHallA:2011ayy, JeffersonLabHallA:2014yxb} Collaborations.
In these experiments, the final-state hadrons are either pions, kaons, and protons, or unidentified charged hadrons.
The phenomenology extractions of the Sivers function based on these data have been performed by many groups~\cite{Anselmino:2005ea,Collins:2005ie, Anselmino:2008sga,Bacchetta:2011gx,Sun:2013hua,Echevarria:2014xaa,Martin:2017yms,Boglione:2018dqd,Echevarria:2020hpy,Bury:2020vhj,Bacchetta:2020gko,Bury:2021sue,Zeng:2022lbo}.
Recently, the COMPASS Collaboration reanalyzed the data collected in 2010 and reported the Sivers asymmetry in $\rho^{0}$ meson production for the first time~\cite{Alexeev:2022wgr}.
According to the factorization theorem, the Sivers function should not depend on the type of hadrons detected in the final state. Therefore the COMPASS measurement can provide a test of the universality of the Sivers function.

Furthermore, the primary decay products of vector mesons are pseudoscalar mesons. For instance, the $\rho^{0}$ meson predominantly decays into two pions, and the $K^{*}$ meson primarily decays into a kaon and a pion. Studying the Sivers asymmetry in vector meson production not only elucidates the Sivers effect in these processes but also deepens our understanding of the Sivers effect in pseudoscalar mesons, such as pions and kaons.
The future facilities, such as the Electron-Ion Collider ({\tt EIC})~\cite{Accardi:2012qut, AbdulKhalek:2021gbh} and the Electron-ion Collider in China ({\tt EicC})~\cite{Anderle:2021wcy}, are expected to yield high-statistics data with broad kinematic coverage, enabling further studies of the Sivers effect.

In this paper, we conduct calculations of the Sivers asymmetry for vector meson production in the SIDIS process within the TMD factorization. The comparison with the COMPASS measurement~\cite{Alexeev:2022wgr} serves as a test of 
the universality of the Sivers function in the SIDIS process for different final state hadrons. Besides, we
also predict the Sivers asymmetries of $\rho^{0}$ and $K^{*}$ productions in {\tt EIC} and {\tt EicC} kinematics based on our current knowledge of the Sivers functions. While different global analyses of the Sivers function~\cite{Bury:2021sue,Zeng:2022lbo} can describe the COMPASS data equally well, the predictions for the Sivers asymmetry with these global analyses at future experiments are differentiable. Therefore, more precise constraints are needed to deepen our understanding of nucleon spin structures.

\section{Theoretical formalism}
\label{sec: formalism}

We consider the SIDIS process with unpolarized lepton beam scattering off a transversely polarized target nucleon, 
\begin{align}
\label{SIDIS process}
        l(\ell)+N(P, S_{\perp}) \longrightarrow l\left(\ell^{\prime}\right)+ h \left(P_h\right)+X,
\end{align}
where $l$ represents the lepton, $N$ represents the target nucleon, $h$ denotes the final-state vector meson $\rho^0$ or $K^*$, and $X$ stands for the undetected hadronic system. The four-momenta of the corresponding particles are labeled in parentheses. $S_{\perp}$ is the transverse polarization of the nucleon.

\begin{figure}[htp]
    \centering
    \includegraphics[width=0.5\textwidth]{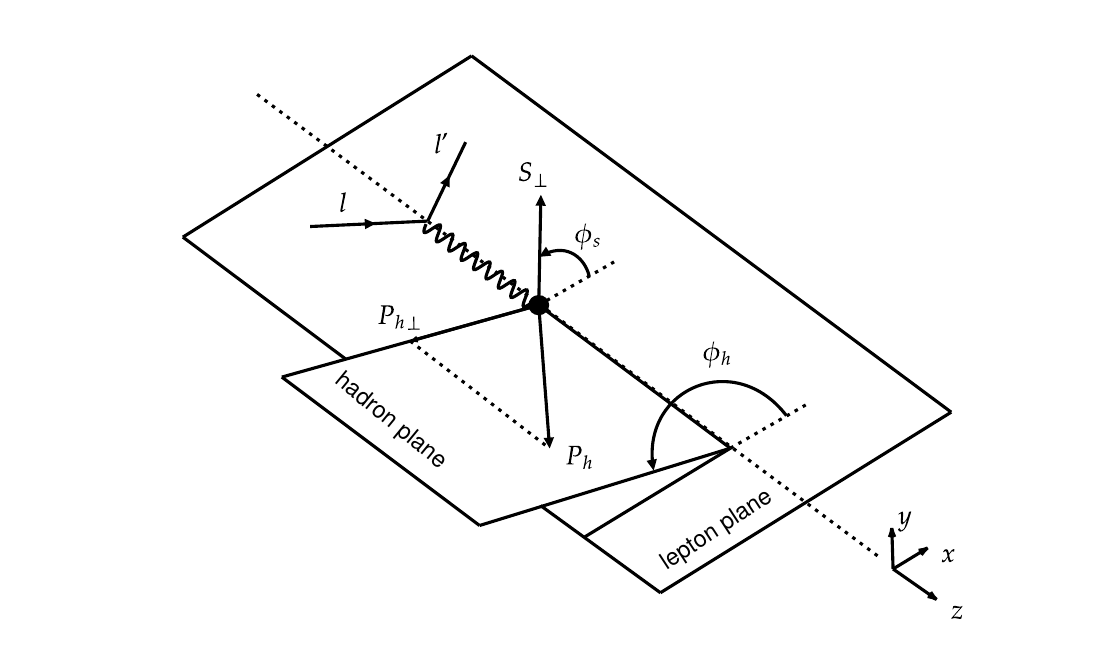}
    \caption{The Trento conventions of SIDIS kinematic variables.}
    \label{fig:coordinate_system}
\end{figure}

Within the one-photon-exchange approximation, the differential cross section is expressed as
\begin{equation}
    \label{eq:cross_section}
    \begin{aligned}
    \frac{\mathrm{d} \sigma}{\mathrm{d}x_{B} \, \mathrm{d}y \, \mathrm{d}z_{h} \, \mathrm{d}P_{h \perp}^{2} \, \mathrm{d}\phi_{h} \, \mathrm{d}\phi_{S}} 
    =&\frac{\alpha^{2}}{x_{B} y Q^{2}} \frac{y^{2}}{2\left(1 - \epsilon\right)} \left(1+\frac{\gamma^{2}}{2x_{B}}\right) 
    \left[F_{UU}(x_{B}, z_{h}, P_{h\perp}, Q^{2})  \right. \\
    & \left. + |S_\perp|\sin\left(\phi_{h} - \phi_S \right)F^{\sin\left(\phi_{h} - \phi_S \right)}_{UT}(x_{B}, z_{h}, P_{h\perp}, Q^{2})  + \ldots\right].
    \end{aligned}
\end{equation}
The kinematic variables in the above equation are defined as
\begin{align}
    x_{B} = \frac{Q^{2}}{2 P \cdot q}, 
    \quad 
    y = \frac{P \cdot q}{P \cdot \ell}, 
    \quad 
    z_{h} = \frac{P \cdot P_{h}}{P \cdot q}, 
    \quad
    \gamma = \frac{2x_{B} M}{Q}, 
    \quad  
    \epsilon = \frac{1-y-\frac{1}{4}\gamma^{2}y^{2}}{1-y+\frac{1}{2}y^{2}+\frac{1}{4}\gamma^{2}y^{2}},
\end{align}
where $M$ is the target nucleon mass, and $q$ is the momentum of the virtual photon with the virtuality $q^2 = - Q^2$.
The azimuthal angles $\phi_h$ and $\phi_S$ are defined in the virtual photon-nucleon frame, as illustrated in Fig.~\ref{fig:coordinate_system}, following the Trento conventions~\cite{Bacchetta:2004jz}. 
In this paper, we investigate the Sivers effect, focusing on two structure functions, $F_{UU}$ and $F_{UT}^{\sin(\phi_{h} - \phi_S)}$. Other structure functions are represented with an ellipsis in Eq.~\eqref{eq:cross_section}. 
The complete formula for the cross section of spin-1 final state hadron can be found in Ref.~\cite{Bacchetta:2000jk}.
The Sivers asymmetry can be related to the structure functions through the following equation,
\begin{align}
    A_{UT}^{\sin(\phi_h - \phi_S)}(x_{B}, z_{h}, P_{h\perp}, Q^{2})
    \equiv \frac{ 2 \int \mathrm{d}\phi_{h} \mathrm{d}\phi_{S} \sin(\phi_{h}-\phi_{S}) \frac{\mathrm{d}\sigma }{\mathrm{d}x_{B} \, \mathrm{d}y \, \mathrm{d}z_{h} \, \mathrm{d}P_{h \perp}^{2} \, \mathrm{d}\phi_{h} \, \mathrm{d}\phi_{S} } }{
        \int \mathrm{d}\phi_{h} \mathrm{d}\phi_{S} \frac{\mathrm{d}\sigma }{\mathrm{d}x_{B} \, \mathrm{d}y \, \mathrm{d}z_{h} \, \mathrm{d}P_{h \perp}^{2} \, \mathrm{d}\phi_{h} \, \mathrm{d}\phi_{S} }
    } 
    = \frac{F_{U T}^{\sin \left(\phi_{h} - \phi_{S}\right)}(x_{B}, z_{h}, P_{h\perp}, Q^{2}) }{F_{U U}(x_{B}, z_{h}, P_{h\perp}, Q^{2})}.
\end{align}

At low transverse momentum, one can apply the TMD factorization to express the structure functions in terms of the corresponding TMD PDFs and TMD FFs. To implement the evolution, it is convenient to perform a transverse Fourier transform,
\begin{align}
    F_{U U}\left(x_{B}, z_{h}, P_{h \perp}, Q\right) 
    &= \int \frac{\mathrm{d}^2 b}{(2 \pi)^2} e^{i \mathbf{q}_{\perp} \cdot \mathbf{b}} 
    \widetilde{F}_{U U}(x_{B}, z_{h}, b, Q), 
    \label{eq: FUU QCD factorization} \\
    F_{\rm Sivers}^\alpha\left(x_{B}, z_{h}, P_{h \perp}, Q\right) 
    &= \int \frac{\mathrm{d}^2 b}{(2 \pi)^2} e^{i \mathbf{q}_{\perp} \cdot \mathbf{b}} 
    \widetilde{F}_{\text {Sivers }}^\alpha(x_{B}, z_{h}, b, Q),
    \label{eq: Sivers QCD factorization}
\end{align}
where ${\bf q}_\perp  = - {\bf P}_{h\perp}/z_{h}$ and $F_{\mathrm{Sivers}}^\alpha$ is related to $F_{UT}^{\sin \left(\phi_{h} - \phi_{S}\right) }$ via 
\begin{align}
    \sin \left(\phi_{h} - \phi_{S}\right) F_{U T}^{\sin \left(\phi_{h} - \phi_{S}\right)} &= 
    \varepsilon_{\alpha \beta \rho \sigma}\frac{P^{\rho} q^{\sigma}S_{\perp}^\alpha}{P \cdot q \sqrt{1+\gamma^{2}}}  F_{\text {Sivers }}^\beta,
\end{align}
with the convention $\varepsilon_{0123} = 1$ for the totally antisymmetric tensor. Then the $b$-space structure functions are expressed with TMD PDFs and TMD FFs as
\begin{align}
\widetilde{F}_{U U}(x_{B}, z_{h}, b, Q) 
&= H\left(\mu, Q\right) \sum_{q}e_{q}^{2} 
\widetilde{f}_{1,q/p}\left(x_{B}, b, \mu, \zeta_1 \right)
\widetilde{D}_{1,h/q}\left(z_{h}, b, \mu, \zeta_2 \right),
\\
\widetilde{F}_{\mathrm{Sivers}}^{\alpha}(x_{B}, z_{h}, b, Q) 
&= H\left(\mu, Q\right) \sum_{q}e_{q}^{2}
\left( -i M b^{\alpha} \right) 
\widetilde{f}_{1T, q/p}^{\perp}\left(x_{B}, b, \mu, \zeta_1 \right) 
\widetilde{D}_{1,h/q}\left(z_{h}, b, \mu, \zeta_2\right),
\end{align}
where $H(\mu,Q)$ is the hard factor, $e_q$ is the fractional electric charge of the struck quark, and the summation runs over all flavors. The $b$-space TMD functions are given by
\begin{align}
f_{1,q/p}\left(x, k_{\perp}^{2}, \mu, \zeta\right) 
&= \int_0^{\infty} \frac{\mathrm{d} \, b}{2\pi}   b J_{0}\left(b k_{\perp}\right) \widetilde{f}_{1,q/p}\left(x, b, \mu, \zeta\right),
\\
D_{1,h/q}\left(z, p_{\perp}^{2}, \mu, \zeta\right) 
&= \int_0^{\infty} \frac{\mathrm{d}\, b}{2\pi} \ b  J_{0}\left(b \frac{p_{\perp}}{z} \right) \widetilde{D}_{1,h/q}\left(z, b, \mu, \zeta\right),
\label{eq:TMDD1k}\\
\frac{k_{\perp}}{M} f_{1 T,q/p}^{\perp}\left(x, k_{\perp}^2, \mu, \zeta\right) 
&= \int_0^{\infty} \frac{\mathrm{d}\, b }{2 \pi} b^2 M J_1\left(b k_{\perp}\right) \widetilde{f}_{1 T,q/p}^{\perp}(x, b, \mu, \zeta),
\end{align}
where $J_{0,1}$ is the Bessel function of the first kind. The $\mu$ and $\zeta$ are the renormalization scale and the rapidity (Collins-Soper) scale.

The evolution of the TMD PDF and TMD FF is governed by the Collins-Soper equation~\cite{Collins:1981uk},
\begin{align}
    \label{eq: TMD evolution equations2}
    \zeta \frac{\mathrm{d} \widetilde{F}(x, b ; \mu, \zeta)}{\mathrm{d} \zeta} & =-\mathcal{D}(\mu, b) \widetilde{F}(x, b ; \mu, \zeta),
\end{align}
and the renormalization group equation,
\begin{align}
    \label{eq: TMD evolution equations1}
    \mu^2 \frac{\mathrm{d} \widetilde{F}(x, b , \mu, \zeta)}{\mathrm{d} \mu^2} & =\frac{\gamma_F(\mu, \zeta)}{2} \widetilde{F}(x, b , \mu, \zeta),
\end{align}
where $\widetilde{F}(x, b, \mu, \zeta)$ stands for any TMD PDF or TMD FF. $\gamma_{F}$ is the TMD anomalous dimension and $\mathcal{D}(\mu, b)$ is the rapidity anomalous dimension (RAD). One can formally solve the equations and obtain the relation,
\begin{align}
\widetilde{F}\left(x, b ; \mu_f, \zeta_f\right)&=\exp \left[\int_P\left(\gamma_F(\mu, \zeta) \frac{d \mu}{\mu}-\mathcal{D}(\mu, b) \frac{d \zeta}{\zeta}\right)\right] \widetilde{F}\left(x, b ; \mu_i, \zeta_i\right) \nonumber \\ &
\equiv R\left[b;\left(\mu_i, \zeta_i\right),\left(\mu_f, \zeta_f\right)\right] \widetilde{F}\left(x, b ; \mu_i, \zeta_i\right).
\end{align}
The evolution factor $R$ is given by a path integral from $\left(\mu_{i}, \zeta_{i}\right)$ to the final point $\left(\mu_{f}, \zeta_{f}\right)$. According to the integrable condition, as can be derived from Eqs.~\eqref{eq: TMD evolution equations1} and \eqref{eq: TMD evolution equations2},
\begin{equation}
    \label{eq: integrability condition}
    \zeta \frac{\mathrm{d}}{\mathrm{d}\zeta} \gamma_{F}\left(\mu, \zeta\right) 
    = - \mu \frac{\mathrm{d}}{\mathrm{d}\mu}\mathcal{D}\left(\mu, b\right),
\end{equation}
the evolution factor is in principle path independent. However, in practical applications one need to truncate the perturbation expansion and then the evolution factor differs from path to path. Here we follow the $\zeta$-prescription~\cite{Scimemi:2018xaf}, in which the path first goes along the null-evolution curve to $\mu=\mu_f$ and then goes along a straight line to $\zeta=\zeta_f$ with $\mu$ fixed. This approach has been adopted in recent global analyses of the Sivers functions~\cite{Bury:2021sue,Zeng:2022lbo}.

\section{Numerical calculation}
\label{sec: numerical}

With the theoretical formalism above, we can express the Sivers asymmetry $A_{UT}^{\sin\left(\phi_{h} - \phi_{S}\right)}$ as \cite{Bury:2021sue,Zeng:2022lbo} 
\begin{align}
    \label{eq: Sivers asymmetry expression}
        A_{UT}^{\sin\left(\phi_{h} - \phi_{S}\right)}\left(x_{B}, z_{h}, P_{h\perp}, Q^{2}\right) 
        =& \frac{
        - M \sum_{q}e_{q}^{2}
        \int_{0}^{\infty} \frac{\mathrm{d} b}{2 \pi} b^{2} J_{1}\left(\frac{bP_{h\perp}}{z_{h}}\right)
        \left(\frac{Q^2}{\zeta_Q(b)}\right)^{-2 \mathcal{D}(b, Q)}
        \widetilde{f}_{1T, q/p}^{\perp}\left(x_{B}, b\right) 
        \widetilde{D}_{1,h/q}\left(z_{h}, b\right)
        }{
        \sum_{q}e_{q}^{2}
        \int_{0}^{\infty} \frac{\mathrm{d} b}{2 \pi} b J_{0}\left(\frac{bP_{h\perp}}{z_{h}}\right)
        \left(\frac{Q^2}{\zeta_Q(b)}\right)^{-2 \mathcal{D}(b, Q)}
        \widetilde{f}_{1,q/p}\left(x_{B}, b \right)
        \widetilde{D}_{1,h/q}\left(z_{h}, b \right)
        }.
\end{align}
where $\widetilde{f}_{1,q/p}\left(x, b \right)$, $\widetilde{D}_{1,h/q}\left(z, b \right)$, and $\widetilde{f}_{1T, q/p}^{\perp}\left(x, b\right)$ are the unpolarized TMD PDF, the unpolarized TMD FF, and the Sivers function in b-space.
As a common choice, we set $\mu=Q$ and the rapidity scale $\zeta = Q^2$. $\zeta_{Q}(b)$ is the $\zeta_{\mu}(b)$ in Eq.~\ref{eq:zeta_expression} with $\mu = Q$.
The details of the TMD evolution can be found in Appendix.~\ref{Appendix_2}.

\begin{figure}[htbp]
    \includegraphics[width = 0.95\textwidth]{ 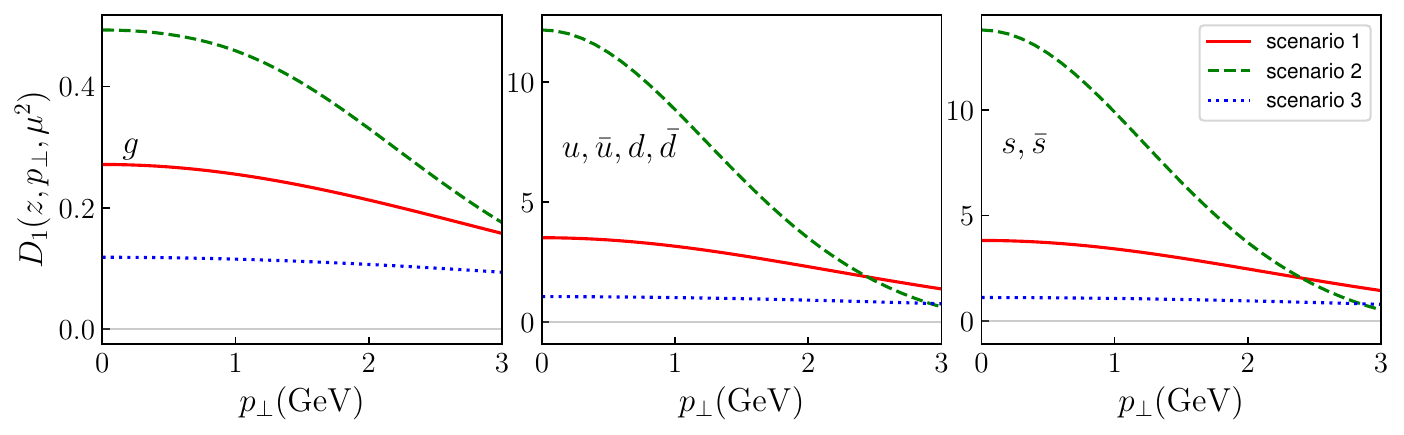}
    \caption{ 
    The fragmentation functions 
    of  $\rho^{0}$ with three scenarios for $D_{\mathrm{NP}}$. The left, central, and right panels show the unpolarized TMD FF of the gluon, valence quarks ($u,\bar{u},d, \bar{d}$), and sea quarks ($s, \bar{s}$) with $z = 0.4,\ \mu^2 = \zeta = 4 \, \mathrm{GeV}^2$. }
    \label{Fig: Three D1NP kperp distribution}
\end{figure}

For unpolarized TMD PDFs of the nucleon, we adopt the SV19 parametrization~\cite{Scimemi:2019cmh}. For unpolarized TMD FFs of the $\rho^0$, there is no available global analysis yet. Here we follow the same parametrization form in Ref.~\cite{Scimemi:2019cmh} for pions as
\begin{align}
    \widetilde{D}_{1,h/f}\left(z, b\right) = & \frac{1}{z^{2}} \sum_{f'} \int_{z}^{1} 
        \frac{\mathrm{d}y}{y} y^{2} \mathbb{C}_{f\rightarrow f'}
        \left(y, b, \mu_{\mathrm{OPE}}^{\mathrm{FF}}\right) 
        D_{1, h/f'}\left(\frac{z}{y}, \mu_{\mathrm{OPE}}^{\mathrm{FF}}\right) 
        D_{\mathrm{NP}}\left(z, b\right).
        \label{eq:TMDD1b}
\end{align}
The coefficient functions $\mathbb{C}_{f\to f'}$ is perturbatively calculable and independent of the hadron type. For the collinear FF $D_{1, h/f'}\left(z, \mu\right)$ for $\rho^0$, we perform a fit using the data generated by Pythia, with details provided in Appendix.~\ref{Appendix_1}. The unpolarized FFs for $K^*$ mesons are calculated following the method in Ref.~\cite{Chen:2020pty}. For the nonperturbative input $D_{\rm NP}(x,b)$, we assume the same form as those for pion in Ref.~\cite{Scimemi:2019cmh},
\begin{align}
    D_{\mathrm{NP}}\left(z, b\right) &= \exp \left[-\frac{\eta_1 z+\eta_2(1-z)}{\sqrt{1+\eta_3(b / z)^2}} \frac{b^2}{z^2}\right]
    \left(1+\eta_4 \frac{b^2}{z^2}\right),
\end{align}
but take three sets of the parameters, as listed in Table~\ref{Tab: parameters of TMD PDF and TMD FF}, for comparison.
Within the formalism presented above, the unpolarized TMD FF in transverse momentum space can be obtained by Eq.~\eqref{eq:TMDD1k}.
Although the unpolarized FF appears in both the numerator and denominator when calculating the Sivers asymmetry, it convolves with different distribution functions, and thus does not cancel out. The different transverse momentum distribution of FFs will affect the magnitude of the Sivers effect.
Since the transverse momentum space is more intuitive than the impact parameter space, we compare these three scenarios of the distribution in $p_{\perp}$ space.
The scenario-1 takes the same parameters as those for pion, and thus the same transverse momentum dependence. The scenario-2 gives more concentrated transverse momentum distributions. The scenario-3 gives more spread transverse momentum distributions. A comparison among the three scenarios is shown in Fig.~\ref{Fig: Three D1NP kperp distribution}.

\begin{table*}[h t b p]
    \caption{The values of the parameters in the parametrization of unpolarized TMD FFs for vector mesons. 
    The units are $\mathrm{GeV}^2$  
    }
    \label{Tab: parameters of TMD PDF and TMD FF}
    \centering
    \begin{tabular}{c c c c c}
    \hline \hline
    scenarios                 &  $\eta_1$      &  $\eta_2$      &  $\eta_3$      &  $\eta_4$  \\
    scenario 1           &  0.260         &  0.476         &  0.478         &  0.483  \\
    scenario 2           & 0.078    & 0.143         & 0.143         & 0.145 \\
    scenario 3           & 0.78          & 1.428         & 1.434         & 1.449 \\
    \hline\hline
    \end{tabular}
\end{table*}

For the Sivers functions, we adopt two recent global analyses, BPV20~\cite{Bury:2021sue} and ZLSZ~\cite{Zeng:2022lbo}. These parametrizations are obtained by analyzing existing Sivers asymmetry measurements for pion and kaon productions.
Both of these parametrizations directly parametrize the initial distribution of the Sivers function in the $b$ space, which is different with the previous method that indirectly parametrized the Sivers function through the Qiu-Sterman function. This parametrization approach is more direct. The evolution formalism chosen by both parametrizations is the method presented above. 
Although ZLSZ parametrization used kaon data in the fitting process, the Sivers functions of $s$ and $\bar{s}$ quarks were set to zero. In the BPV20 parametrization, the contributions from $s$ and $\bar{s}$ Sivers functions are retained.
Considering the universality of the Sivers function in the SIDIS process for different final state hadrons, they should be able to describe the $\rho^0$ meson production data.
In the following calculations, the precision we use is NLO+NNLL, the same as ZLSZ~\cite{Zeng:2022lbo}.

\begin{figure}[h t b p]
    \centering
        \includegraphics[width = 0.9\textwidth]{ 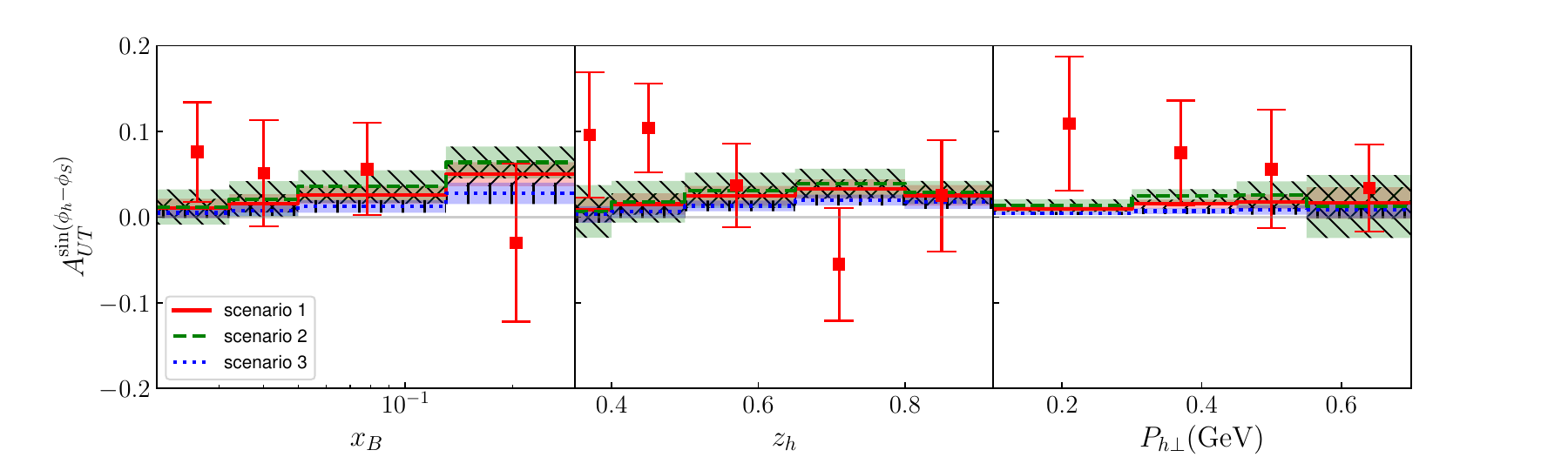}
        \includegraphics[width = 0.9\textwidth]{ 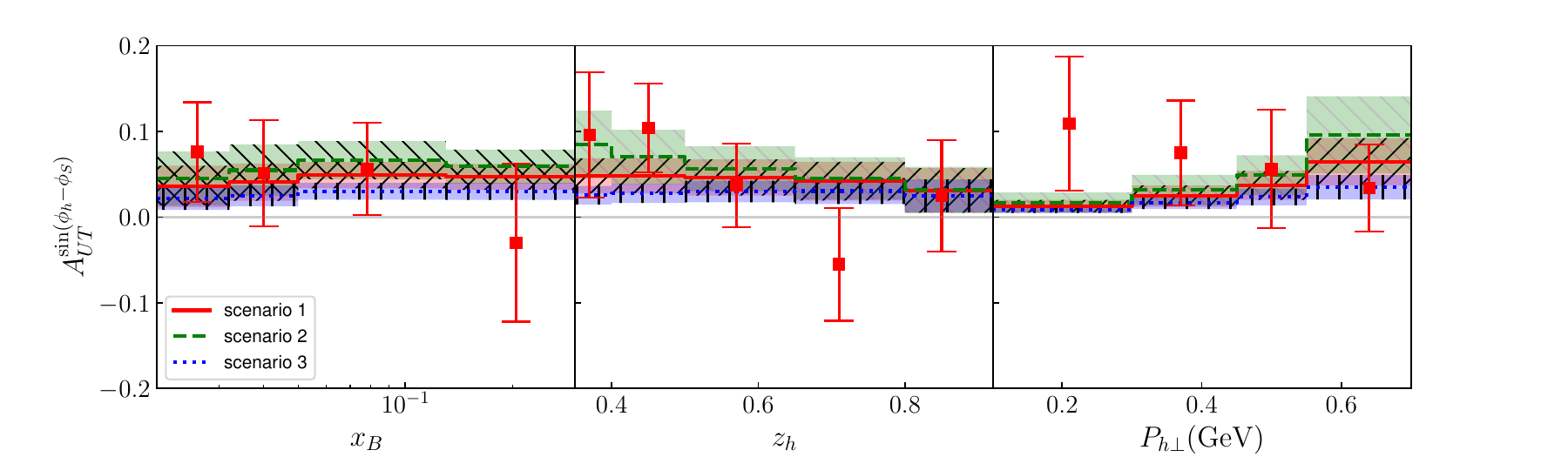}
    \caption{ The Sivers asymmetry of $\rho^0$ production in comparison with the COMPASS measurement~\cite{Alexeev:2022wgr}. 
    The theoretical bands in upper panels are evaluated using the ZLSZ parametrization~\cite{Zeng:2022lbo}, and those in lower panels are evaluated using BPV20 parametrization~\cite{Bury:2021sue}. The bands represent the standard deviation evaluated via the replicas given by the fits of the Sivers functions.
    The three scenarios correspond to different assumptions of the transverse momentum dependence of the TMD FFs for $\rho^0$.}
    \label{Fig: COMPASS Sivers}
\end{figure}

In Fig.~\ref{Fig: COMPASS Sivers}, we compare the theoretical predictions with the COMPASS data~\cite{Alexeev:2022wgr}. The uncertainty bands are evaluated from the Monte Carlo replicas of the fits of the Sivers functions. One can observe that both ZLSZ and BPV20 fit predict positive Sivers asymmetries, but they exhibit different trends, particularly in their dependence on $z_h$. Although one might expect the $z_h$ dependence to be primarily driven by the FFs, it actually reflects the influence of the chosen Sivers function through the convolution. The Sivers function shapes the transverse momentum weighting, ultimately determining the $z_h$ dependence. Additionally, BPV20 predicts a larger asymmetry than ZLSZ as $P_{h\perp}$ increases. Although such different features are obtained from the two parametrizations, one can observe that both can describe the COMPASS data well. Hence the universality of the Sivers function in the SIDIS process for different final state hadrons is supported within the current precision of data. Comparing the three scenarios of the transverse momentum dependence, one can find that the scenario-2 gives greater Sivers asymmetry, because it gives more concentrated distributions in the transverse momentum. Again, the current data are insufficient to distinguish between these scenarios. Therefore, high precision data in future experiments are desired.

\begin{figure}[h t b p]
    \centering
    \includegraphics[width=1\textwidth]{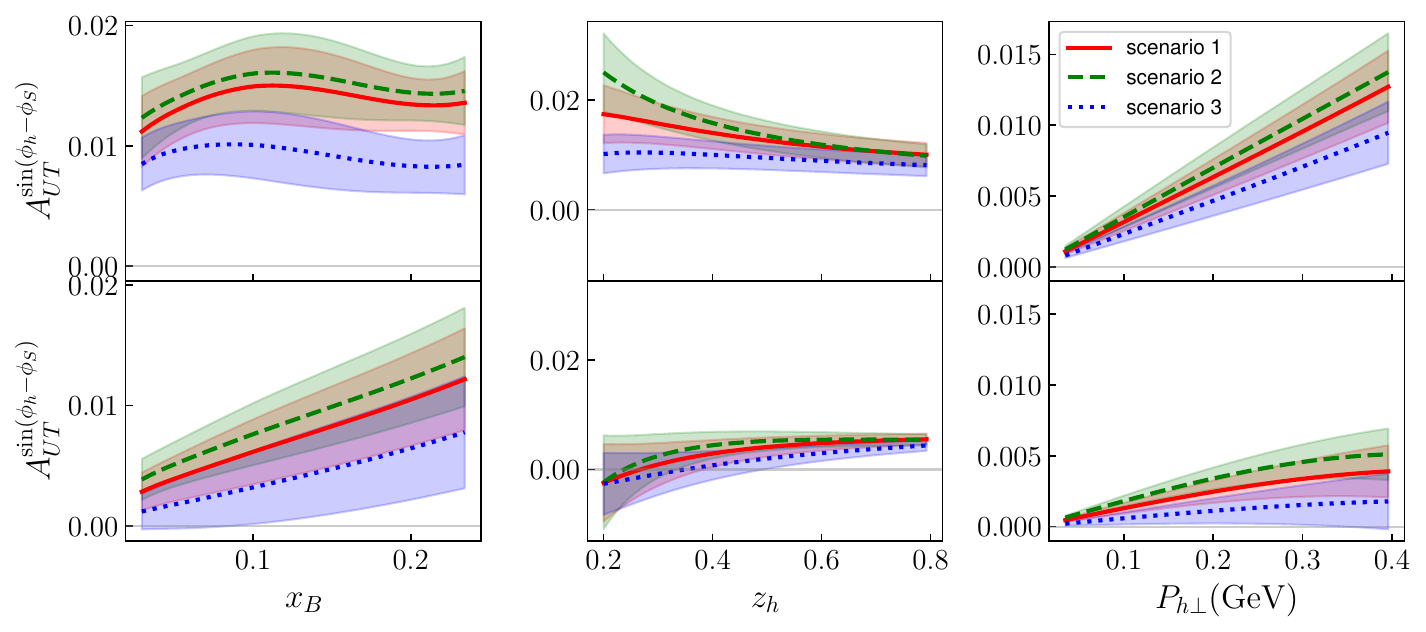}
    \caption{The Sivers asymmetry of $\rho^{0}$ meson at the EIC. The kinematics are set at $\sqrt{s} = 100\,\rm GeV$, $y=0.2$, $P_{h\perp} = 0.4\,\rm GeV$ for $x_B$ and $z_h$ panels, $z_h = 0.48$ for $x_B$ and $P_{h\perp}$ panels, and $x_B=0.05$ for $z_h$ and $P_{h\perp}$ panels. The upper panels are evaluated using BPV20 parametrization, and the lower panels are evaluated using ZLSZ parametrization.
    }
    \label{fig: expectation_EIC_rho0}
\end{figure}

\begin{figure}[h t b p]
    \centering
    \includegraphics[width=1\textwidth]{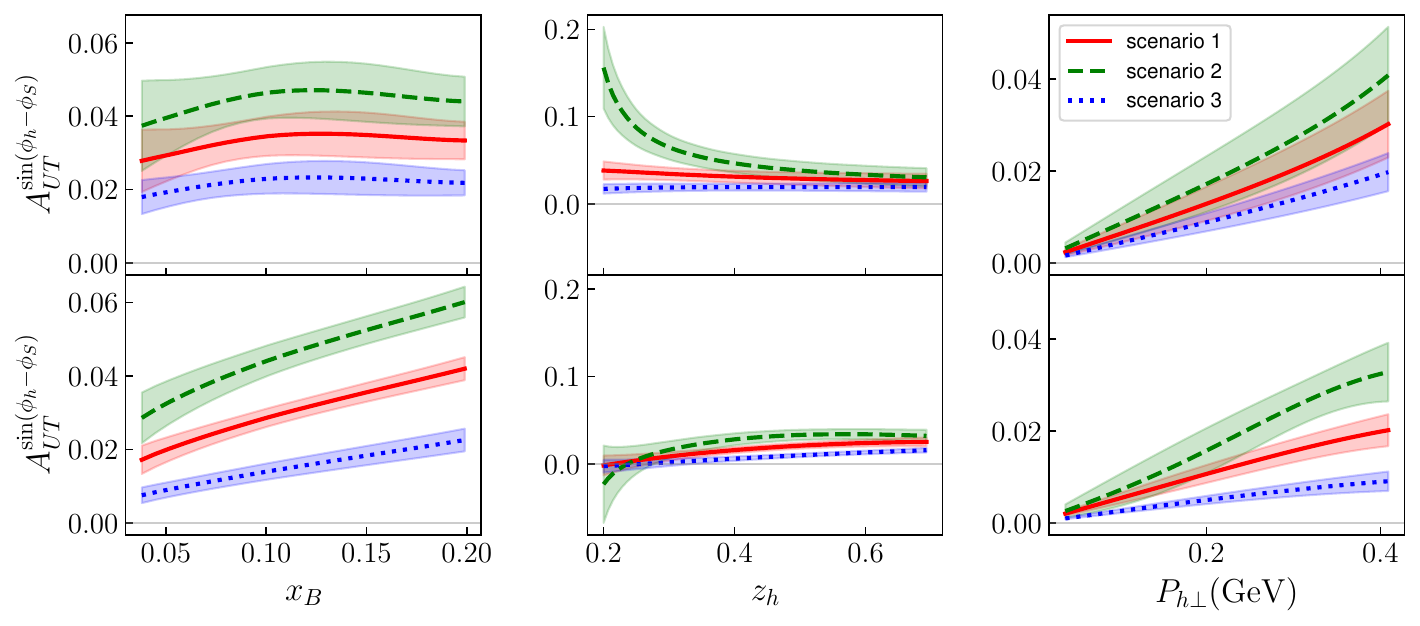}
    \caption{The Sivers asymmetry of $\rho^{0}$ meson at the EicC. The kinematics are set at $\sqrt{s} = 16.7\,\rm GeV$, $y=0.2$, $P_{h\perp} = 0.4\,\rm GeV$ for $x_B$ and $z_h$ panels, $z_h = 0.48$ for $x_B$ and $P_{h\perp}$ panels, and $x_B=0.05$ for $z_h$ and $P_{h\perp}$ panels. The upper panels are evaluated using BPV20 parametrization, and the lower panels are evaluated using ZLSZ parametrization.
    }
    \label{fig: expectation_EicC_rho0}
\end{figure}

In addition to the comparison with existing COMPASS data, we also make predictions at {\tt EIC} and {\tt EicC} kinematics. The predictions of the Sivers asymmetry of $\rho^0$ meson production at {\tt EIC} are shown in Fig.~\ref{fig: expectation_EIC_rho0}, and those at {\tt EicC} are shown in Fig.~\ref{fig: expectation_EicC_rho0}. 
One can observe that the Sivers asymmetries are predominantly positive, and the trends for both parametrizations of the Sivers function align with those at the kinematic region of COMPASS. 
Comparing the three scenarios reveals that a more concentrated $p_{\perp}$ distribution of the TMD FFs leads to larger Sivers asymmetries. 
Comparing the predictions in Fig.~\ref{fig: expectation_EIC_rho0} and Fig.~\ref{fig: expectation_EicC_rho0}, one can find that more significant Sivers effects are expected at the EicC kinematics than those at the EIC kinematics, although the shapes of the curves are very similar. This feature is expected since the distributions at a higher scale is more diluted by the shower characterized by the evolution.
Comparing the results from the two parametrizations, one can observe that BPV20 and ZLSZ provide very different predictions. As both parametrizations are determined by fitting existing data, such different predictions indicate that future EicC and EIC will provide constraints on the extraction of the Sivers functions and precise tests of their universality for different final state hadrons.

\begin{figure}[h t b p]
    \centering    
    \includegraphics[width=0.9\textwidth]{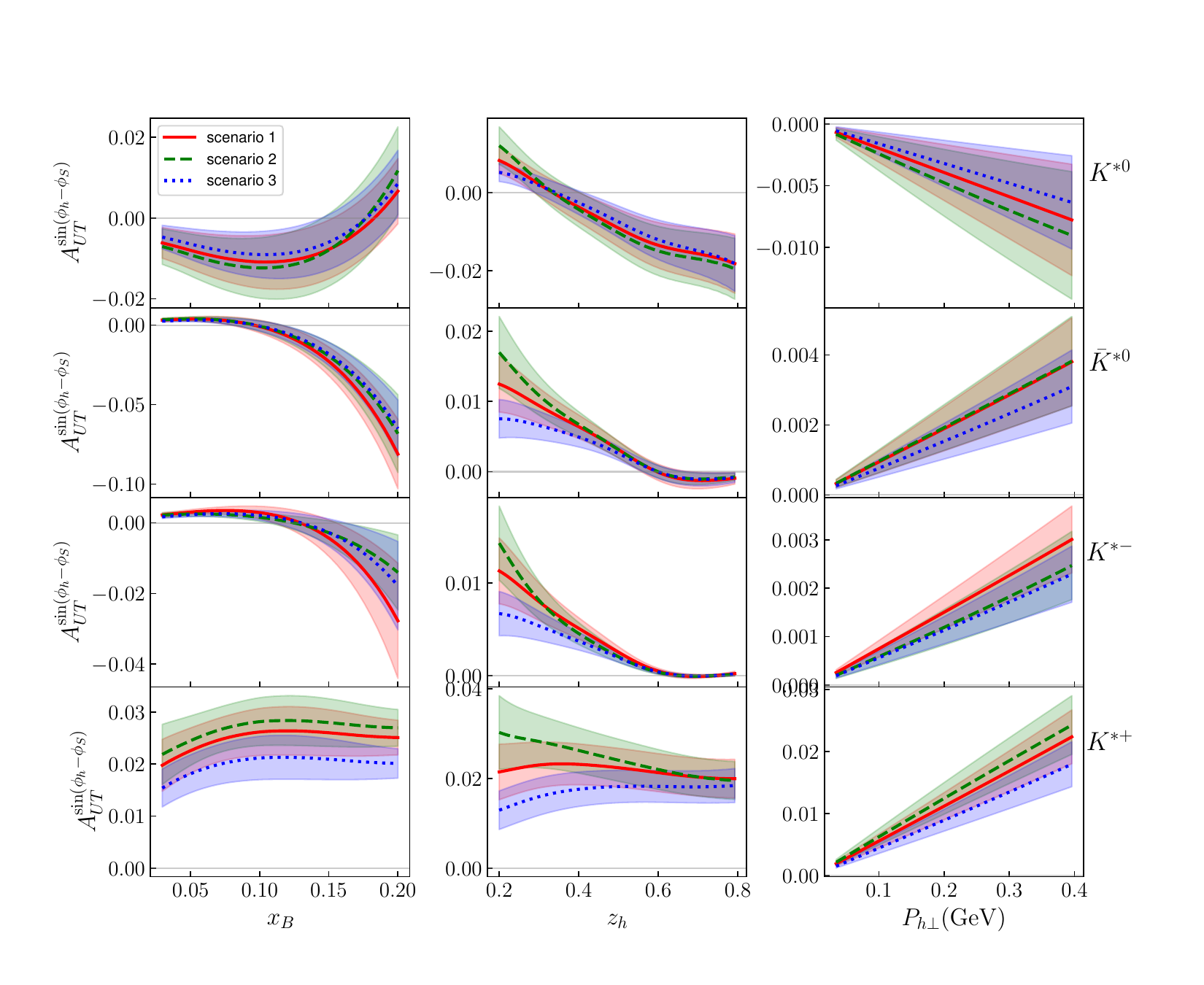}
    \caption{The Sivers asymmetry of $K^{*}$ meson productions evaluated using BPV20 parametrization at the EIC kinematics.}
    \label{fig: Kstar_EIC_BPV20}
\end{figure}

\begin{figure}[h t b p]
    \centering    
    \includegraphics[width=0.9\textwidth]{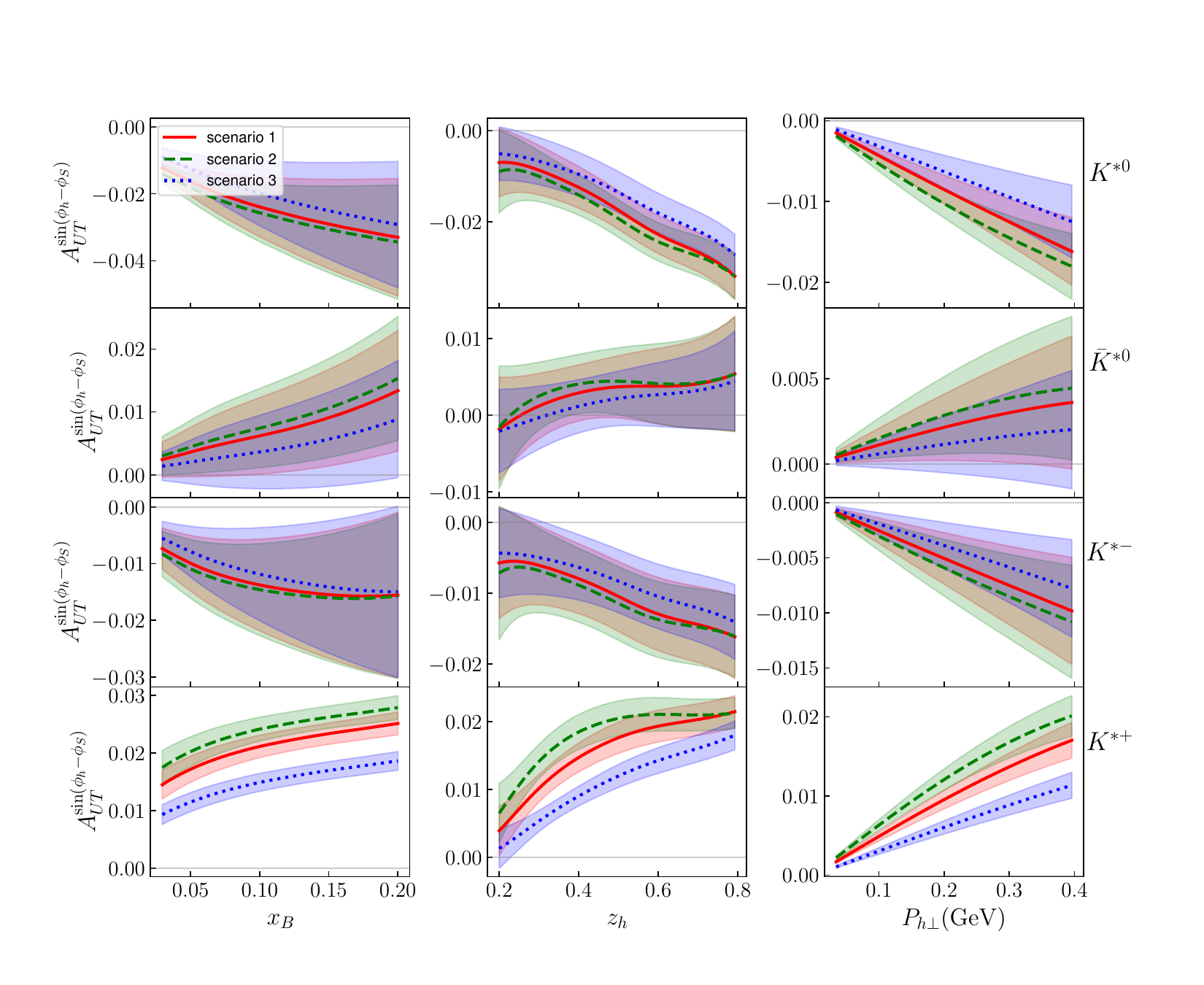}
    \caption{The Sivers asymmetry of $K^{*}$ meson productions evaluated using ZLSZ parametrization at the EIC kinematics.}
    \label{fig: Kstar_EIC_ZLSZ}
\end{figure}

\begin{figure}[h t b p]
    \centering    
    \includegraphics[width=0.9\textwidth]{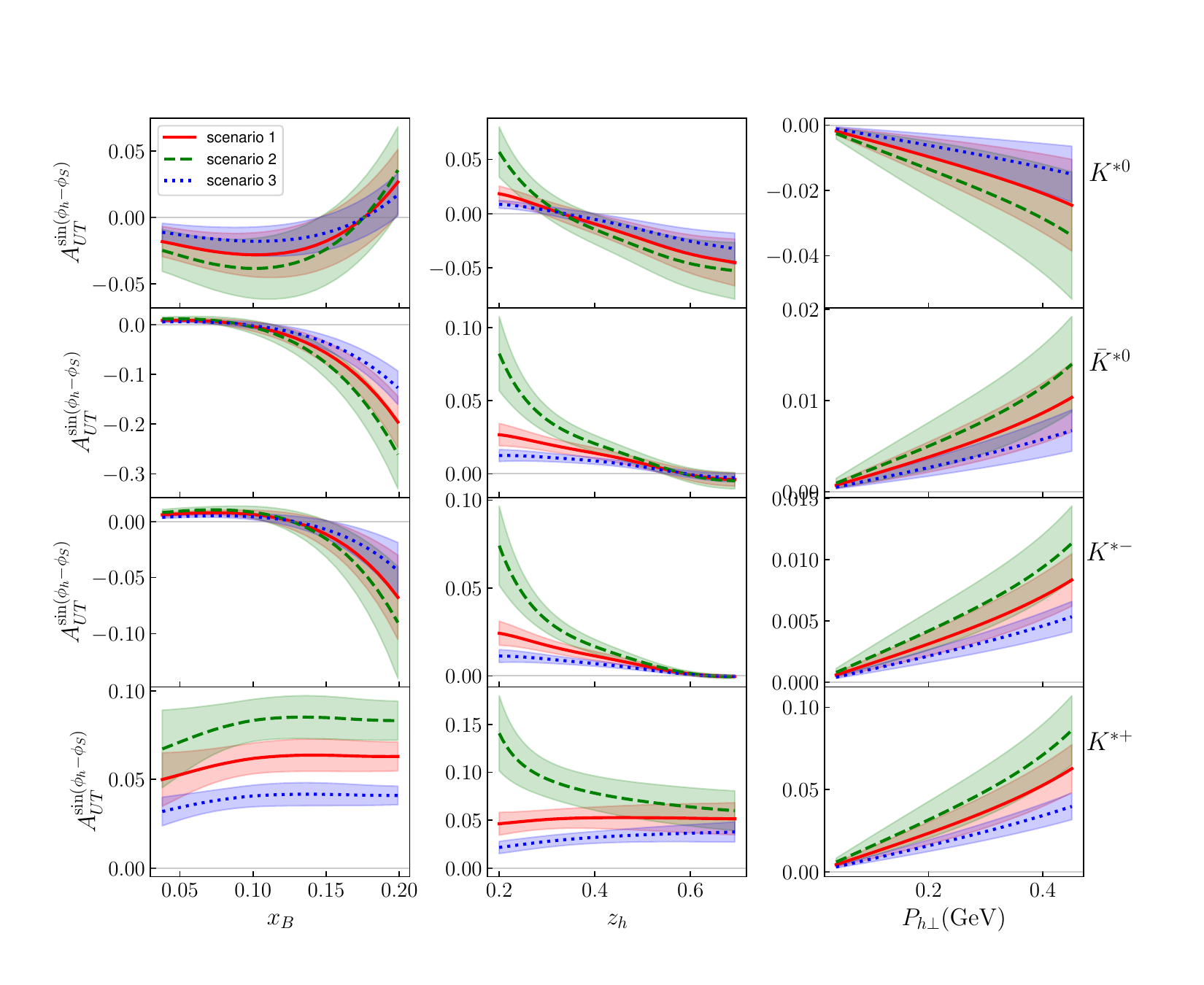}
    \caption{The Sivers asymmetry of $K^{*}$ meson productions evaluated using BPV20 parametrization at the EicC kinematics.}
    \label{fig: Kstar_EicC_BPV20}
\end{figure}

\begin{figure}[h t b p]
    \centering 
    \includegraphics[width=0.9\textwidth]{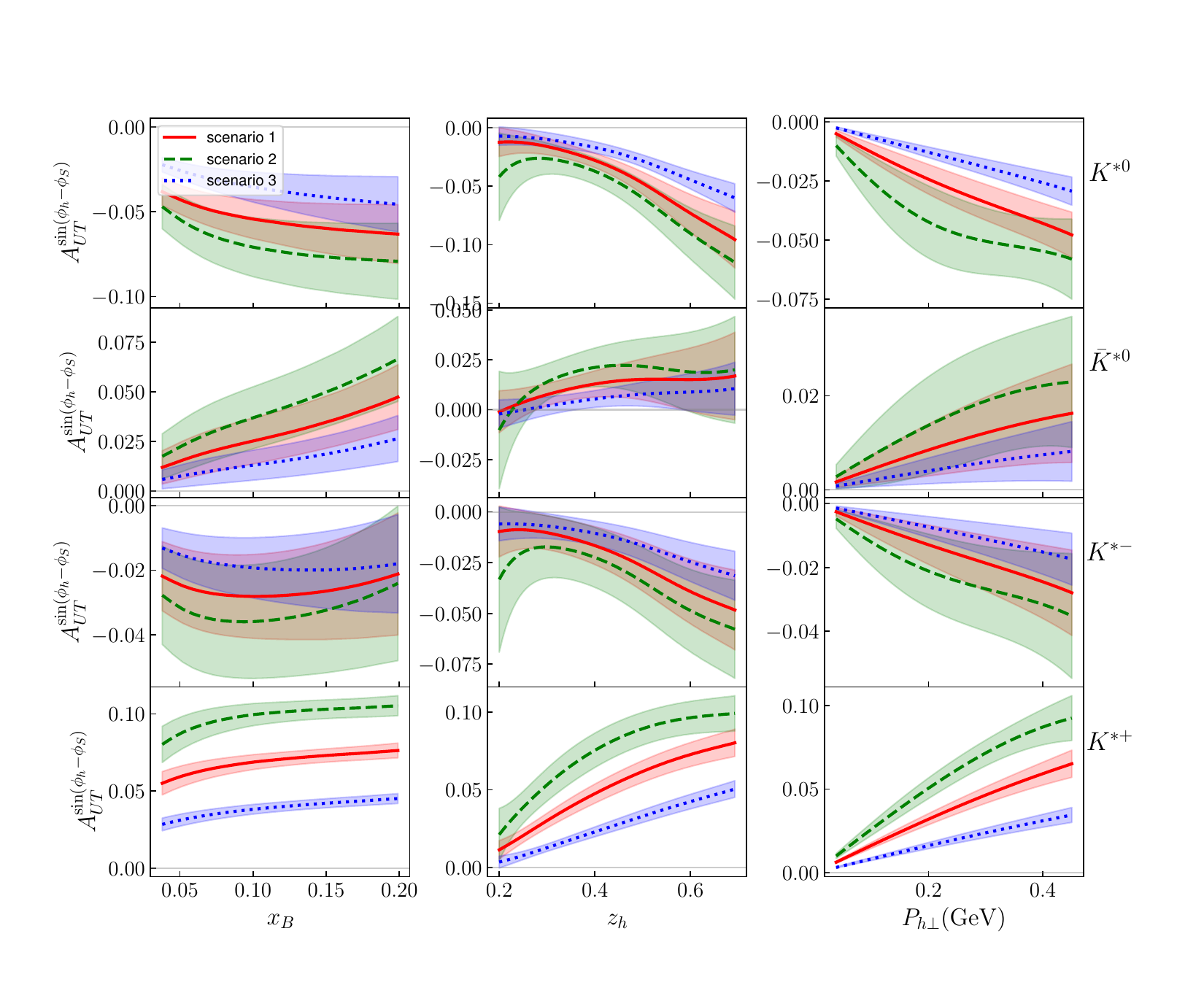}
    \caption{The Sivers asymmetry of $K^{*}$ meson productions evaluated using ZLSZ parametrization at the EicC kinematics.}
    \label{fig: Kstar_EicC_ZLSZ}
\end{figure}

We also calculate the Sivers asymmetries for $K^{*}$ mesons productions, including $K^{*+}$, $K^{*-}$, $K^{*0}$, and $\bar{K}^{*0}$, as shown in Figs.~\ref{fig: Kstar_EIC_BPV20} and~\ref{fig: Kstar_EIC_ZLSZ} at the EIC kinematics, and in Figs.~\ref{fig: Kstar_EicC_BPV20} and~\ref{fig: Kstar_EicC_ZLSZ} at the EicC kinematics. 
In the $x_{B}$-panel of the second and third rows of Fig.~\ref{fig: Kstar_EIC_BPV20}, we observe a sign change in the Sivers asymmetry as it crosses $x_{B} = 0.1$. Since in the $z_{h}$-panel we use $x_{B} = 0.05$, the sign at $z_{h} = 0.48$ in the $z_{h}$-panel and at $P_{h\perp} = 0.4 \, \mathrm{GeV}$ in the $P_{h\perp}$-panel is 
the same as the sign at $x_{B} = 0.05$ in the $x_{B}$-panel. 
A similar situation also occurs in Fig.~\ref{fig: Kstar_EicC_BPV20}.
Similar to the $\rho^{0}$ meson production, a more concentrated $p_{\perp}$ distribution of the unpolarized TMD FFs gives greater Sivers asymmetries in the $K^{*}$ meson productions. 
One can also observe that the Sivers asymmetry predicted by ZLSZ parametrization is more sensitive to the $p_{\perp}$ distribution of the unpolarized TMD FF, which makes the curves of the three scenarios more separated than those from the BPV20 predictions. 
The asymmetry values at the EicC kinematics are in general greater than those at the EIC kinematics region. With these aspects, the measurements at EicC are expected to have strict constraints if assuming equal statistics as EIC. 
The two different parametrizations of the Sivers function yield significant differences in the Sivers effect. For instance, under the BPV20 parametrization, the $K^{*0}$ production shows a decreasing dependence on $P_h$, which differs from the increasing trend observed for the other three $K^*$ mesons. On the other hand, under the ZLSZ parametrization, $K^{*0}$ and $ K^{-}$ also exhibit a decreasing trend. The $x_B$ and $z_h$ dependencies also differ significantly between the two parametrizations.
So the Sivers effect of $K^{*}$ meson productions in the SIDIS process will refine the determination of the Sivers functions.


\section{Summary}
\label{sec: summary}

We investigate the Sivers asymmetry $A_{UT}^{\sin\left(\phi_{h} - \phi_{S}\right)}$ of vector meson productions in SIDIS process within the TMD factorization. The Sivers asymmetry arises from the convolution of the Sivers functions of the nucleon and the unpolarized FFs of the final-state hadron. 

In the numerical calculation, we use two parametrizations, BPV20 and ZLSZ, for the Sivers functions. Three scenarios are assumed for the transverse momentum dependence of the FFs in order to investigate the influence of the Sivers asymmetry. We find both parametrizations of the Sivers functions equipped with various scenarios of the TMD FFs can well describe the COMPASS data. This agreement supports the universality of the Sivers functions in SIDIS for different final state hadron, which are extracted from data on pion and kaon productions mesons. However, the large uncertainties of current data cannot distinguish different fits of Sivers functions, even if noticeable  different behaviors are obtained. 
Therefore, more precise data from future experiments are needed.

In addition, we make predictions of the Sivers asymmetries of vector meson productions, including $\rho^0$ and $K^*$, at the EIC and EicC kinematics based on our current knowledge of the Sivers functions. The results show predominantly positive asymmetry values. For the $K^{*}$ mesons, significant difference is observed between the predictions from two parametrizations.
The expectations of Sivers asymmetry for $K^{*0}$, $\bar{K}^{*0}$, and $K^{*-}$ exhibit variations in sign in the $x_{B}$ panels. The expectations of Sivers asymmetry for  $K^{*-}$ exhibit variations in sign in the $P_{h\perp}$ panels.
A more concentrated unpolarized TMD FF distribution corresponds to a more significant Sivers effect for the $\rho^{0}$ meson and $K^{*}$ mesons.
The difference between the BPV20 and ZLSZ parametrizations offers an opportunity for further testing the universality for different final state hadron and improving the determination of the Sivers functions.
Next-generation colliders are expected to provide high-statistics data, thereby advancing our understanding of the nucleon spin structure.

\acknowledgments{
We thank Peng Sun, Zhe Zhang, Jing Zhao, and Xiaoyan Zhao for useful discussions. This work was supported by the National Key R\&D Program of China, No.~2024YFA1611004, and by the National Natural Science Foundation of China (Grants No.~12175117, No.~12475084, and No.~12321005) and Shandong Province Natural Science Foundation (Grants No.~ZFJH202303 and No.~ZR2024MA012).
}

\appendix
\section{ The rapidity anomalous dimension and the rapidity scale}
\label{Appendix_2}
In the perturbative region where $1/b \gg \Lambda_{QCD}$, one can expand $\mathcal{D}\left(\mu, b\right)$ in powers of the strong coupling constant as, 
\begin{align}
    \label{eq: D_pert expansion}
    \mathcal{D}_{\text{pert}}\left(\mu, b\right) = \sum_{n = 0}^{\infty} a_{s}^{n} 
    d_{n}\left(\mathbf{L}_{\mu}\right),
\end{align}
where $a_{s} = \alpha_{s}/(4\pi)$, the subscription ``$\text{pert}$'' indicates this formula is only valid in perturbative region, and $\mathbf{L}_{\mu}$ can be written as 
\begin{align}
    \mathbf{L}_{\mu} = \ln\left( \frac{\mu^{2} b^{2}}{4e^{-2\gamma_{E}}} \right),
\end{align}
with $\gamma_E \approx 0.577$ being the Euler-Mascheroni constant.
Up to two-loop order, the expansion coefficients $d_{n}\left(\mathbf{L}_{\mu}\right)$ are 
\begin{align}
        d_0\left(\mathbf{L}_\mu\right) & =0, \\
        d_1\left(\mathbf{L}_\mu\right) & =\frac{\Gamma_0}{2} \mathbf{L}_\mu, \\
        d_2\left(\mathbf{L}_\mu\right) & =\frac{\Gamma_0}{4} \beta_0 \mathbf{L}_\mu^2
        +\frac{\Gamma_1}{2} \mathbf{L}_\mu+d_2(0),
\end{align}
where 
\begin{align}
    d_2(0)=C_F C_A\left(\frac{404}{27}-14 \zeta_3\right)-\frac{112}{27} T_R N_f C_F ,
\end{align}
with $\zeta_{3} = 1.202$, $C_{F} = 4/3$, $C_{A} = 3$ and $T_{R} = 1/2$.
$\Gamma_{n}$ and $\beta_{n}$ have the following expressions:
\begin{align}
    \Gamma_0 =&  4 C_F ,\\
    \Gamma_1 =&  4 C_F\left[\left(\frac{67}{9}-\frac{\pi^2}{3}\right) 
                C_A-\frac{20}{9} T_R N_f\right],\\
    \Gamma_2= & 4 C_F\left[C_A^2\left(\frac{245}{6}-\frac{134 \pi^2}{27}
                +\frac{11 \pi^4}{45}+\frac{22}{3} \zeta_3\right) \right. 
               +C_A T_F n_f\left(-\frac{418}{27}+\frac{40 \pi^2}{27}
                -\frac{56}{3} \zeta_3\right) \nonumber\\
              & \left.+C_F T_F n_f\left(-\frac{55}{3}+16 \zeta_3\right)
                -\frac{16}{27} T_F^2 n_f^2\right], \\
    \beta_0 = &\frac{11}{3} C_A-\frac{4}{3} T_F n_f , \\
    \beta_1 = &\frac{34}{3} C_A^2-\frac{20}{3} C_A T_F n_f-4 C_F T_F n_f , \\
    \beta_2 = & \frac{2857}{54} C_A^3+\left(2 C_F^2-\frac{205}{9} C_F C_A 
               -\frac{1415}{27} C_A^2\right) T_F n_f  
               +\left(\frac{44}{9} C_F+\frac{158}{27} C_A\right) T_F^2 n_f^2. 
\end{align}
The expansion \eqref{eq: D_pert expansion} has a small convergence radius, as the expansion variable $\left(a_s \mathbf{L}_{\mu}\right)$ rapidly exceeds 1 with increasing $b$. To improve the convergence properties of RAD, we use the resummed expression~\cite{Scimemi:2018xaf, Bizon:2018foh, Echevarria:2012pw}, which can be written as 
\begin{equation}
\begin{aligned}
    \mathcal{D}_{\mathrm{resum }}(\mu, b) =& -\frac{\Gamma_0}{2 \beta_0} \ln (1-X)  +\frac{a_s}{2 \beta_0(1-X)}\left[-\frac{\beta_1 \Gamma_0}{\beta_0}(\ln (1-X)+X)+\Gamma_1 X\right] \\
    & +\frac{a_s^2}{(1-X)^2}\left[\frac{\Gamma_0 \beta_1^2}{4 \beta_0^3}\left(\ln ^2(1-X)-X^2\right)\right.  +\frac{\beta_1 \Gamma_1}{4 \beta_0^2}\left(X^2-2 X-2 \ln (1-X)\right) \\
    & +\frac{\Gamma_0 \beta_2}{4 \beta_0^2} X^2-\frac{\Gamma_2}{4 \beta_0} X(X-2)  \left.+C_F C_A\left(\frac{404}{27}-14 \zeta_3\right)-\frac{112}{27} T_R N_f C_F\right].
\end{aligned}
\end{equation}

As we mentioned above, the expansion for $\mathcal{D}\left(\mu, b\right)$, Eq.\eqref{eq: D_pert expansion}, is only valid in the region where $b \gtrsim 1/\Lambda_{\mathrm{QCD}}$. In order to obtain the RAD in the entire $b$ region, we can model it as follows
\begin{align}
    \label{eq: full RAD}
    \mathcal{D}(\mu, b)=\mathcal{D}_{\mathrm{resum }}\left(b_{*}, \mu\right)+d_{\mathrm{NP}}(b),
\end{align}
where $b_{*}$ is defined as 
\begin{align}
    b_{*} = \frac{b}{\sqrt{1+\frac{b^2}{B_{\mathrm{NP}}^2}}},
\end{align}
which allows for a smooth transition from the perturbative region to the non-perturbative region.
The function $d_{\mathrm{NP}}\left(b\right)$ is used to describe the large $b$ behavior of $\mathcal{D}\left(\mu, b\right)$, and we adopt the form in Ref.~\cite{Bertone:2019nxa},
\begin{align}
    d_{\mathrm{NP}}\left(b\right) = c_{0} b b_{*},
\end{align}
which is linear at the large-$b$ region as suggested in Refs.~\cite{Tafat:2001in, Vladimirov:2020umg, Hautmann:2020cyp, Scimemi:2019cmh}.
The parameters $B_{\mathrm{NP}} = 1.93 \, \mathrm{GeV}^{-1}$, $c_{0} = 0.0391 \, \mathrm{GeV}^{2}$ can be found in Ref.~\cite{Scimemi:2019cmh}.

In the $\zeta$-prescription, the null-evolution curves are governed by the equation
\begin{equation}
    \label{Eq: null-evolution line}
    \frac{\mathrm{d} \ln \zeta_{\mu}\left( b\right)}{\mathrm{d}\ln\mu^2} =
    \frac{\gamma_F(\mu, \zeta_{\mu}\left(b\right))}{2 \mathcal{D}\left(\mu, b\right)}
\end{equation}
and the saddle point is the solution with boundary conditions \cite{Scimemi:2018xaf}
\begin{equation}
    \mathcal{D}\left(\mu_0, b\right) = 0, \quad \gamma_{F}\left(\mu_{0}, \zeta_{\mu_{0}}(b)\right) = 0.
\end{equation}

Using Eq.~\eqref{eq: full RAD} as an input to solve Eq.~\eqref{Eq: null-evolution line}, one can obtain a solution that is independent of the specific form of the $\mathcal{D}(\mu, b)$ as 
\begin{align}
    \zeta_\mu^{\mathrm{exact}}(b)=\mu^2 e^{-g(\mu, b) / \mathcal{D}(\mu, b)},
\end{align}
where the expression of $g\left(\mu, b\right)$ up to two-loop order can be written as 
\begin{equation}
\begin{aligned}
    g(\mu, b) =& \frac{1}{a_s} \frac{\Gamma_0}{2 \beta_0^2}\left\{e^{-p}-1+p+
        a_s\left[\frac{\beta_1}{\beta_0}\left(e^{-p}-1+p-\frac{p^2}{2}\right)\right.-\frac{\Gamma_1}{\Gamma_0}\left(e^{-p}-1+p\right)+
        \frac{\beta_0 \gamma_1}{\Gamma_0} p\right] \\ &
        +a_s^2\left[\left(\frac{\Gamma_1^2}{\Gamma_0^2}-
        \frac{\Gamma_2}{\Gamma_0}\right)(\cosh p-1) \left.+\left(\frac{\beta_1 \Gamma_1}{\beta_0 \Gamma_0}
        -\frac{\beta_2}{\beta_0}\right)(\sinh p-p)+\left(\frac{\beta_0 \gamma_2}{\Gamma_0}
        -\frac{\beta_0 \gamma_1 \Gamma_1}{\Gamma_0^2}\right)
        \left(e^p-1\right)\right]\right\},
\end{aligned}
\end{equation}
with 
\begin{align}
    p =& \frac{2\beta_{0}\mathcal{D}\left(\mu, b\right)}{\Gamma_{0}}, \\
    \gamma_1 =& -6 C_F, \\
    \gamma_2 =& C_F^2\left(-3+4 \pi^2-48 \zeta_3\right) 
                 +C_F C_A\left(-\frac{961}{27}-\frac{11 \pi^2}{3}+52 \zeta_3\right) 
                 +C_F T_R N_f\left(\frac{260}{27}+\frac{4 \pi^2}{3}\right). 
\end{align}
At extremely small $b$ region, $\zeta_\mu^{\mathrm{exact }}$ should be matched to $\zeta_{\mu}^{pert}\left(\mu,b\right)$, which is the result by using $\mathcal{D}_{\mathrm{pert}}\left(\mu, b\right)$ as an input to solve Eq.~\eqref{Eq: null-evolution line},
\begin{align}
    \zeta_\mu^{\text {pert }}(b)=\frac{2 \mu e^{-\gamma_E}}{b} e^{-v(\mu, b)},
\end{align}
where the expression of $v\left(\mu, b\right)$ up to two-loop order can be written as 
\begin{align}
    v(\mu, b)=\frac{\gamma_1}{\Gamma_0}+a_s\left[\frac{\beta_0}{12} 
    \mathbf{L}_\mu^2+\frac{\gamma_2+d_2(0)}{\Gamma_0}-
    \frac{\gamma_1 \Gamma_1}{\Gamma_0^2}\right].
\end{align}
However, there could be numerical difficulties to match because it is very difficult to obtain exact numerical cancellation of perturbative series of logarithms with $\zeta_{\mu}^{\mathrm{exact}}$. 
To by-pass this problem, we follow the approach in Ref.~\cite{Scimemi:2019cmh} which express $\zeta_{\mu}\left(b\right)$ as 
\begin{align}
    \label{eq:zeta_expression}
        \zeta_\mu( b)= & \zeta_\mu^{\mathrm{pert}}( b) e^{-b^2 / B_{\mathrm{NP}}^2} 
         +\zeta_\mu^{\mathrm{exact }}( b)\left(1-e^{-b^2 / B_{\mathrm{NP}}^2}\right). 
\end{align}
With this expression, one can use $\zeta_{\mu}^{\mathrm{pert}}\left(b\right)$ which cancels the logarithm exactly at small $b$, and use $\zeta_{\mu}^{\mathrm{exact}}\left(b\right)$ at large $b$.

\section{Global fit of \texorpdfstring{$D_{1,\rho^{0}/f}\left(z, \mu\right)$}{}}
\label{Appendix_1}
In this appendix, we obtain the unpolarized parton to $\rho^{0}$ fragmentation functions by performing a global fit of $\rho^0$ production data of Pythia.

The observable that is used to perform global fit is $F^{h}(z, Q^{2})$. It is defined by the hadron-production cross section and the total hadronic cross section~\cite{Altarelli:1979kv, Nason:1993xx, Furmanski:1981cw},
\begin{equation}
    \begin{aligned}
        F^{h}\left(z,Q^{2}\right)=&\frac{1}{\sigma_{\mathrm{tot }}}\frac{d\sigma\left(e^{+}e^{-}\rightarrow hX\right)}{dz} = \frac{1}{\sum_{q}\hat{e}_{q}^{2}}\left[2F_{1}^{h}\left(z,Q^{2}\right)+F_{L}^{h}\left(z,Q^{2}\right)\right],
    \end{aligned}
    \label{Fragmentation function}
\end{equation}
where $Q^{2}$ is the virtual photon or $Z^0$ momentum squared in $e^{+}e^{-}\rightarrow\gamma,Z$. The variable $z$ is defined by 
\begin{equation}
    z\equiv\frac{E_{h}}{\sqrt{s}/2}=\frac{2E_{h}}{Q},
\end{equation}
where $E_{h}$ and $\sqrt{s}/2$ are the hadron and beam energies, respectively. 
\begin{equation}
    \sigma_{\mathrm{tot}}=\sum_{q} \hat{e}_q^2 \frac{4\pi \alpha_{em}^2(Q^2) }{s} \left[1+\frac{\alpha_{s}\left(Q^{2}\right)}{\pi}\right]
\end{equation}
is the total cross section for $e^{+}e^{-}\to hadrons$, where the perturbative correction is included up to the NLO. 
$\hat{e}_{q}$ is the electroweak charge which is given by
\begin{align}
    \hat{e}_q^2=e_q^2-2 e_q \chi_1\left(Q^2\right) V_e V_q+\chi_2\left(Q^2\right)\left(1+V_e^2\right)\left(1+V_q^2\right), 
\end{align}
where 
\begin{align}        
    \chi_1(s) &= \frac{1}{16 \sin ^2 \Theta_W \cos ^2 \Theta_W} \frac{s\left(s-M_Z^2\right)}{\left(s-M_Z^2\right)^2+\Gamma_Z^2 M_Z^2}, \\
    \chi_2(s) &= \frac{1}{256 \sin ^4 \Theta_W \cos ^4 \Theta_W} \frac{s^2}{\left(s-M_Z^2\right)^2+\Gamma_Z^2 M_Z^2}.
\end{align}
Here, $M_{Z}$ and $\Gamma_{Z}$ are the mass and decay width of $Z$ boson, respectively. 
The other electroweak couplings are given in terms of the Weinberg angle $\Theta_{W}$ by
\begin{align}
    V_e &= -1+4 \sin ^2 \Theta_W,\\
    V_u &= +1-\frac{8}{3} \sin ^2 \Theta_W,\\
    V_d &= -1+\frac{4}{3} \sin ^2 \Theta_W.
\end{align}

To NLO accuracy the unpolarized ``time-like" structure functions $F_{1}^{h}$ and $F_{L}^{h}$ in Eq.~\eqref{Fragmentation function} are given by 
\begin{align}
        &2 F_{1}^{h}\left(z, Q^{2}\right)= \sum_{q} \hat{e}_{q}^{2}\left\{\left[D_{h/q}\left(z, Q^{2}\right)+D_{h/\bar{q}}\left(z, Q^{2}\right)\right]\right. 
        \nonumber\\
        &+\frac{\alpha_{s}\left(Q^{2}\right)}{2 \pi}\left. \left[C_{q}^{1} 
        \otimes\left(D_{h/q}+D_{h/\bar{q}}\right)+C_{g}^{1} \otimes D_{h/g}\right]\left(z, Q^{2}\right)\right\},
        \\
        &F_{L}^{h}\left(z, Q^{2}\right)= \frac{\alpha_{s}\left(Q^{2}\right)}{2 \pi} \sum_{q} \hat{e}_{q}^{2}\left[C_{q}^{L} 
        \otimes\left(D_{h/q}+D_{h/\bar{q}}\right) \right. \nonumber \\
        & \left. + C_{g}^{L} \otimes D_{h/g} \right] \left(z, Q^{2}\right).
\end{align}
The relevant NLO coefficient functions $C_{q,g}^{1,L}$ in the $\overline{\mathrm{MS}}$ scheme can be found in Appendix.A of~\cite{deFlorian:1997zj}.

The fragmentation functions are expressed in terms of a number of parameters at the initial scale $Q_{0}^{2} = 2.4 \, \mathrm{GeV^2}$. Since they should be zero at $z = 1$, a simple polynomial form is taken
\begin{equation}
    D_{h/i}\left(z, Q_{0}^{2}\right)=N_{i}^{h} z^{\alpha_{i}^{h}}(1-z)^{\beta_{i}^{h}}, \quad \left(i = u, d, s, g, \bar{u}, \bar{d}, \bar{s}\right),
\end{equation}
where $N_{i}^{h}$, $\alpha_{i}^{h}$, and $\beta_{i}^{h}$ are parameters to be determined by a global $\chi^{2}$ analysis.

Considering the constituent quark composition $\rho^{0} = \frac{1}{\sqrt{2}}(u\bar{u}-d\bar{d})$ and the charge symmetry, we take the same favored fragmentation functions for $\rho^{0}$ from $u$, $\bar{u}$, $d$, and $\bar{d}$ quarks:
\begin{equation}
    \begin{aligned}
        D_{\rho^0/u}(z) =&  D_{\rho^0/d}(z) =D_{\rho^0/\bar{u}}(z) =D_{\rho^0/\bar{d}}(z) \\
        =& N_1 \times z^{\alpha_1} \times (1-z)^{\beta_1}.
    \end{aligned}
\end{equation}
The $\rho^{0}$ production from $s$ and $\bar{s}$ are disfavored processes, and they are considered the same at the initial scale:
\begin{equation}
    D_{\rho^0/s}(z) =D_{\rho^0/\bar{s}}(z)= N_2 \times z^{\alpha_2} \times (1-z)^{\beta_2}.
\end{equation}
In addition, a fragmentation function from gluon for $\rho^{0}$ is given by 
\begin{equation}
    D_{\rho^0/g}(z)= N_g \times z^{\alpha_g} \times (1-z)^{\beta_g}.
\end{equation}
So the following parameters are used for the $\rho^{0}$ in our global analysis:
\begin{equation}
    \label{Pars}
    N_{1}^{\rho^0}, \alpha_{1}^{\rho^0}, \beta_{1}^{\rho^0}, \quad N_{2}^{\rho^0}, \alpha_{2}^{\rho^0}, \beta_{2}^{\rho^0}, \quad N_{g}^{\rho^0}, \alpha_{g}^{\rho^0}, \beta_{g}^{\rho^0}. 
\end{equation}
We obtain these parameters by performing a global fit of $\rho^{0}$ production data of Pythia. 
The formula that relates $F^{h}\left(z, Q^{2}\right)$ to the data from Pythia is given by
\begin{equation}
    \begin{aligned}
        F^{\rho^{0}}\left(z, Q^{2}\right)=&\frac{1}{\sigma_{\mathrm{tot}}} \frac{d \sigma\left(e^{+} e^{-} \rightarrow \rho^{0} X\right)}{d z} 
        = \frac{1}{N_{\mathrm{tot}}} \frac{\Delta N\left(e^{+} e^{-} \rightarrow \rho^{0} X\right)}{\Delta z},    
    \end{aligned}
 \end{equation}
where $N_{\mathrm{tot}}$ respects the  number of $e^{+}e^{-} \to \left(\gamma, Z \right) \to q\bar{q}$, $\Delta N\left(e^{+} e^{-} \rightarrow \rho^{0} X\right)$ is the number of $e^{+} e^{-} \rightarrow \rho^{0} X$ with energy fraction $z \in \left[z, z + \Delta z \right)$, and $\Delta z$ is the bin spacing.
Since we only consider $u$, $d$, and $s$ quarks and the corresponding antiquarks, we set the parameters of Pythia so that the $\gamma$ or $Z$ decays only to $u\bar{u}$, $d\bar{d}$ and $s\bar{s}$. 
The total $\chi^{2}$ is calculated by 
\begin{equation}
    \chi^{2} = \sum_{j} \frac{\left(F_{j}^{\text {data }}-F_{j}^{t h e o}\right)^{2}}{\left(\sigma_{j}^{\text {data }}\right)^{2}},
\end{equation}
where $F_{j}^{\mathrm{data}}$ and $F_{j}^{\mathrm{theo}}$ are Pythia and theoretical values of $F^{\rho^{0}}\left(z,Q^{2}\right)$, respectively. The errors $\sigma_{j}^{\mathrm{data}}$ are only calculated from statistical errors $\sigma_{j}^{\mathrm{stat}}$. The assigned parameters are determined so as to obtain the minimum $\chi^{2}$. The optimization of the functions is done by the MINUIT2~\cite{Hatlo:2005cj}.
The QCD evolution is performed by the package QCDNUM~\cite{Botje:2010ay}.
The values of the parameters $N$, $\alpha$, and $\beta$ in Eq.~\eqref{Pars} resulting from the NLO fit are collected in Table~\ref{Tab:fit result}.

\begin{table*}[h t b p]
        \caption[]{Parameters determined for $\rho^{0}$.}
        \label{Tab:fit result}
        \centering
        \begin{tabular}{c c c c}
            \hline
            \hline
            $\chi^{2}/d.o.f.$ = 749/743\\
            \hline
            function           & $N$                   & $\alpha$               & $\beta$                \\
            $D_{\rho^0/u}$     & 0.4224 $\pm$ 0.0013   & -0.6119 $\pm$ 0.0035   & 1.2448 $\pm$ 0.0037    \\
            $D_{\rho^0/s}$     & 0.3346 $\pm$ 0.0044   & -0.7777 $\pm$ 0.0127   & 2.0681 $\pm$ 0.0229    \\
            $D_{\rho^0/g}$     & 129.04 $\pm$ 21.159   &  3.2234 $\pm$ 1.6011   & 9.9508 $\pm$ 8.5609    \\
            \hline
            \hline
        \end{tabular}
\end{table*}

\end{document}